# Localized and itinerant dichotomy of electrons in Ba(Fe,Co)$_2$As$_2$

H. Q. Yuan[1,*], L. Jiao[1], F. F. Balakirev[2], J. Singleton[2], C. Setty[3], J. P. Hu[3], T. Shang[1], L. J. Li[1], G. H. Cao[1], Z. A. Xu[1], B. Shen[4], H. H. Wen[4]

[1]*Department of Physics, Zhejiang University, Hangzhou, Zhejiang 310027, China*
[2]*NHMFL, Los Alamos National Laboratory, MS E536, Los Alamos, NM 87545, USA.*
[3]*Department of Physics, Purdue University, West Lafayette, IN 47907, USA*
[4]*Beijing National Laboratory for Condensed Matter Physics, Institute of Physics, Chinese Academy of Science, Beijing 10080, China.*
*Email: hqyuan@zju.edu.cn

## Abstract

Variant approaches, either based on the Fermi surface nesting or started from the proximity to a Mott-insulator, were proposed to elucidate the physics in iron pnictides, but no consensus has been reached. A fundamental problem concerns the nature of their 3d electrons. Here we report the magnetoresistivity ($\rho_{xx}$) and the Hall resistivity ($\rho_{xy}$) of Ba(Fe$_{1-x}$Co$_x$)$_2$As$_2$ (x=0 and 0.05) in a magnetic field of up to 55T. The magnetic transition is extremely robust against magnetic field, giving strong evidence that the magnetic ordering is formed by local moments. The magnetic state is featured with a huge magnetoresistance and a distinguished Hall resistivity, $\rho_{xy}(\mu_0 H)$, which shows a pronounced parabolic field dependence, while the paramagnetic state shows little magnetoresistance and follows a simple linear magnetic field dependence on the Hall resistivity. Analyses of our data, based on a two-carrier model, demonstrate that the electron carriers in the magnetic state rapidly increase upon applying a magnetic field, partially compensating the loss of electron carriers at $T_M$. We argue that the 3d-electrons in **Ba(Fe$_{1-x}$Co$_x$)$_2$As$_2$ are divided into those who are close to forming localized moments controlling the magnetic transition and the others giving rise to complex transport properties through their interaction with the former.**

### Introduction

Dichotomy of localized/itinerant characters of electrons has been demonstrated in some correlated electron systems. For example, in the heavy fermion compounds CeRhIn$_5$ and UPd$_2$Al$_3$[1,2], it was shown that the magnetic order originates from the localized 4f-electrons but superconductivity develops from the attractions of "heavy" itinerant electrons. Reminiscent of the heavy fermion compounds[3] and the high T$_c$ cuprates[4], superconductivity recently discovered in the iron pnictides seems to be tied up with magnetism as well[5-13]. Intensive efforts have been devoted to the study of magnetism and superconductivity in these compounds, but their origins remain a puzzle. It was initially proposed that the magnetic order in iron pnictides is of spin-density-wave (SDW) type[6,14,15], driven by Fermi-surface nesting. Such a scenario was supported by some angle-resolved photoemission spectroscopy (ARPES) [16] and optical measurements[6], in which an energy gap was observed below T$_M$. In contrast, a strong-coupling approach was also suggested to describe the physics of iron pnictides[17-22], in which the magnetic ordering stems from the superexchange interactions of nearly localized 3d-electrons. Early measurements of quantum oscillations[23], Hall resistivity[11,24] and ARPES[25] in the 122-type compounds have shown a rather low charge carrier density in the magnetic state, with a possible Fermi surface reconstruction at the magnetic/structural transition temperature (T$_M$). The strong Fermi surface reconstruction, together with the reduction of charge carriers at T$_M$, could be attributed to electronic localization as well as a gap opening induced by magnetic ordering.

In an itinerant electron system with a spin density wave transition, a strong external magnetic field may significantly affect the SDW transition. The robustness of a magnetic transition to external magnetic field can be used to check the degree of electronic itinerancy. Thus, strong pulsed magnetic fields, reaching an energy scale similar to the magnetic transition temperature $T_M$, may provide evidence to the nature of the 3d electrons and, therefore, probe the mechanism of magnetism and superconductivity in iron pnictides.

**Experimental results**

Figure 1 shows the magnetoresistivity, $\Delta\rho_{xx}(\mu_0 H)/\rho_0$, of BaFe$_2$As$_2$ at various temperatures. We find that the magnetoresistivity is nearly zero at temperatures above $T_M$ ($T_M \approx 142K$, see below). In contrast, a huge magnetoresistivity is observed in the magnetic state, and its value depends on the orientation of the applied magnetic field. For instance, the magnetoresistivity at 4K reaches 300% and 120% upon applying a field of 50T along the c-axis and the ab-plane, respectively.

To demonstrate the effect of magnetic field on the magnetic transition, the electrical resistivity $\rho_{xx}(T)$ of both doped and undoped samples is plotted as a function of temperature in Fig.2, from which one can observe the following important points: (i) At zero field, a resistive anomaly is observed at $T_M \approx 142K$ and 77K for BaFe$_2$As$_2$ and Ba(Fe$_{0.95}$Co$_{0.05}$)$_2$As$_2$ respectively, consistent with reports in the literature [10,11]. (ii) The magnetoresistivity $\Delta\rho_{xx}(T)/\rho_0$ (see the insets) monotonically decreases with increasing temperature and vanishes at $T_M$. This fact indicates that the resistive anomaly at $T_M$ is unlikely driven by the structural distortion, but rather by magnetic interaction, because the former usually causes little magnetoresistance[26]. Thus we call the transition at $T_M$ a magnetic transition in this context. (iii) The magnetic transition at $T_M$ is nearly independent of the external magnetic field up to 55T or even higher in both the parent and the (under)doped compounds, in disfavour of a conventional SDW-type magnetic state. (iv) A field induced metal-insulator-like crossover occurs near 20 T in the parent compound at low temperatures, irrespective of the orientation of the magnetic field. Together with the observation of the insulator-like resistivity in $Ba(Fe_{0.95}Co_{0.05})_2As_2$ below $T_{M,}$ it might suggest the importance of electronic localization in $BaFe_2As_2$.

In a conventional SDW state, a sufficiently large magnetic field would suppress the energy gap and, therefore, lead to a decrease of the magnetic transition temperature. For example, the spin-flip transition ($T_{sf}$=123K, which is close to $T_M$≈142K in $BaFe_2As_2$) in chromium, an archetypal material for studying the SDW state, is quadratically suppressed by applying a magnetic field[27]. In contrast, the magnetic transition is extremely robust against the magnetic field of over 55T in $Ba(Fe_{1-x}Co_x)_2As_2$. It is noted that the energy scale of magnetic excitations at a field of 55T is compatible with that of the thermal excitations at $T_M$ ($T_M$≈77K for x=0.05) and the spin gap (~10meV for x=0) resolved by neutron scattering[28]. Thus the magnetic transition is expected to be largely suppressed at such a high magnetic field in the conventional SDW scenario. On the other hand, in the strong-coupling local-moment picture the energy scale required to suppress the magnetic transition is of the order of 200meV in terms of the zone-boundary spin-wave energy of $CaFe_2As_2$[29], which is far beyond the magnetic field we applied. Our experimental findings, therefore, provide evidence that the magnetically ordered state in iron pnictides may not be a conventional SDW state, but possibly originates from local moments and is thus rather robust against magnetic field as observed in other strongly correlated materials[30].

The field induced crossover from metallic to insulator-like behaviour in $BaFe_2As_2$, as demonstrated in the low temperature electrical resistivity $\rho_{xx}(T)$, indicates that the charge carriers and their mobility might be strongly affected by magnetic field. Such a point is further supported by simultaneous measurements of the Hall resistivity. In Fig.

3 the Hall resistivity, $\rho_{xy}(\mu_0 H)$, is plotted as a function of magnetic field for (a) $BaFe_2As_2$ and (b) $Ba(Fe_{0.95}Co_{0.05})_2As_2$. One can see that the Hall resistivity has a dramatic change at $T_M$ in $BaFe_2As_2$, but it is weakened in the doped case. In the parent compound, the Hall resistivity follows rather linear field dependence at temperatures above $T_M$, but shows a parabolic-like field dependence upon entering the magnetic state ($T<T_M$). Such behaviour becomes more pronounced with decreasing temperature, resulting in a change of both the Hall resistivity and its slopes from negative to positive at sufficiently large magnetic fields. Note that the minimum in $\rho_{xy}(\mu_0 H)$ occurs at approximately the fields at which the resistivity (see Fig. 2) changes from metallic to insulating. It is difficult to understand such unique behaviour in terms of conventional theories of Hall effect, even after considering possible multi-band effect. As shown below as a prospective scenario, one actually finds that the charge-carrier characteristics may depend on the magnetic field. While the Hall resistivity in $Ba(Fe_{0.95}Co_{0.05})_2As_2$ also deviates from the linear field dependence below $T_M$, it shows a strictly monotonic decrease with increasing magnetic field up to 50T.

The inset of Fig. 3a shows the temperature dependence of the Hall coefficient $R_H(T)$ ($R_H=\rho_{xy}/\mu_0 H$) at various magnetic fields for $BaFe_2As_2$, from which one can see that $R_H(T)$ undergoes a sharp decrease at $T_M$. In small fields, $R_H(T)$ continues to decrease with further cooling down, suggesting a loss of electron carriers in the magnetically ordered state. These findings are in consistence with the measurements of quantum oscillations[23] and the recent ARPES experiments[25] on this series of compounds, which demonstrated that the Fermi surface in the magnetic state only occupies a small fraction of the paramagnetic Brillouin zone. The hump observed in $R_H(T)$ around 80K might be related to the metal-insulator-like crossover as seen in the resistivity $\rho_{xx}(T)$. Indeed, it is difficult to reconcile the striking increase of the Hall coefficient $R_H$s with increasing magnetic field, implying an increase in carrier density, and the small volume of Fermi surface. Similar behaviour is also found in $Ba(Fe_{0.95}Co_{0.05})_2As_2$.

**Discussion**

It is nonetheless worthwhile to assume the iron pnictides to possess a multi-band electronic structure, consisting of at least two hole pockets and two electron pockets[15]. The unusual field dependence of the transverse Hall resistivity and the longitude magnetoresistivity are, therefore, analyzed in terms of an effective two-carrier model[31]:

$$\rho_{xx}(H) = \frac{\rho_e \rho_h (\rho_e + \rho_h) + (\rho_h R_e^2 + R_h^2 \rho_e)(\mu_0 H)^2}{(\rho_e + \rho_h)^2 + (R_e + R_h)^2 (\mu_0 H)^2} \tag{1}$$

$$\rho_{xy}(H) = \frac{(R_e \rho_h^2 + R_h \rho_e^2) + R_e R_h (R_h + R_e)(\mu_0 H)^2}{(\rho_e + \rho_h)^2 + (R_e + R_h)^2 (\mu_0 H)^2} \mu_0 H \tag{2}$$

where $R_e$ ($R_h$) and $\rho_e$ ($\rho_h$) are the Hall coefficient and the electrical resistivity contributed by electron (hole) pockets, respectively.

From Eq. (2), one expects a linear field dependence of the Hall resistivity in the case of equal carrier numbers ($R_e + R_h = 0$) for holes and electrons. Such behaviour of the Hall resistivity is indeed observed in $BaFe_2As_2$ at temperatures above $T_M$, indicating a good balance of electrons and holes in the paramagnetic state. The remarkably distinct behaviour of $\rho_{xy}(\mu_0 H)$ observed below $T_M$, therefore, suggests, in this approach, a dramatic change of the charge carriers in the magnetic state. It is noted that the data of $\rho_{xx}(\mu_0 H)$ and $\rho_{xy}(\mu_0 H)$ at $T < T_M$ cannot be described in terms of the above two-carrier model with constant parameters of $R_e$, $R_h$, $\rho_e$ and $\rho_h$. Following the procedures described in the section of Methods, we can actually determine the field dependence of $R_e$, $R_h$, $\rho_e$ and $\rho_h$ from $\rho_{xx}(\mu_0 H)$ and $\rho_{xy}(\mu_0 H)$. The derived parameters for the parent compound are plotted as a function of magnetic field in Fig. 4, which reproduce well the results of $\rho_{xx}(\mu_0 H)$ and $\rho_{xy}(\mu_0 H)$. Indeed, the charge carriers and their contribution to resistivity are found to depend strongly on the magnetic field. In particular, the Hall coefficient $R_e(\mu_0 H)$ rapidly increases upon applying a magnetic field and then eventually gets saturated, further supporting a field-induced enhancement of the electron carrier density.

On the other hand, the hole-carrier density is eventually reduced in a magnetic field, leading to an increase of the hole resistivity. However, the absolute value of $R_e$ remains larger than that of $R_h$ even in a high magnetic field, suggesting that the application of a magnetic field can only partially compensate the loss of electron carriers at $T_M$. This is again compatible with the argument that portion of the 3d-electrons are localized at $T_M$ which is extremely robust against the external magnetic field. One notes that the same analyses give meaningless parameters for the underdoped compound (x=0.05), likely attributed to the break-up of Fermi liquid behaviour as a result of enhanced spin fluctuations.

Based on the above experimental findings and the fact that the density of states at the Fermi level is dominantly contributed by 3d-electrons[15], we propose the following scenario for the electronic state in iron pnictides: a portion of the 3d-electrons become localized at $T_M$ and form a long-range magnetically ordered state, which is rather robust against the magnetic field; the other 3d-electrons are itinerant, giving rise to complex transport properties and even superconductivity as observed in the underdoped compound. The hybridization between the localized electrons and the itinerant electrons might open an energy gap at the Fermi surface, as observed in some measurements of spectroscopies[6, 28]. Application of external magnetic field could suppress the spin gap, giving rise to an increase of the electron carriers. The complex field dependence of the itinerant electrons and holes results in very complicated transport properties as reflected in Hall resistivity. These salient findings essentially unveil the nature of the magnetic state in the parent compounds, and the evidence of electronic dichotomy is fundamental for investigating the enigmatic properties of the iron pnictides, including the relation between magnetism and superconductivity in general.

Note that such a scenario of localized and itinerant dichotomy of electrons in iron pnictides is further strengthened by the recent work on the $(Tl,K)Fe_xSe_2$, in which superconductivity appears on the boarder of an AFM Mott insulator and the hole pockets seem to be absent [13,32]. These new observations disagree with the Fermi surface nesting of itinerant electrons and provide more direct evidence of local moment antiferromagnetism in $(Tl,K)Fe_xSe_2$.

**Methods.**

The formulas given by Eq.(1) and Eq.(2) are derived by assuming that there are two uncorrelated charge carriers whose densities and mobilities are not affected by applying external magnetic field. For a system in which the physical properties of charge carriers may strongly depend on the external magnetic field, we may still make use of the above formulas to extract the field dependence of the parameters, $R_e$, $R_h$, $\rho_e$ and $\rho_h$. In the following, we describe a method of data analysis that naturally extracts these parameters without making any additional assumption.

We notice that the following relation can be derived from Eq.(1) and Eq.(2):

$$\mu(H) = \frac{R_e + R_h}{\rho_e + \rho_h} = -\frac{R_{xy}(H) - R_{xy}(0)}{\rho_{xx}(H) - \rho_{xx}(0)} \tag{3}$$

Here, $R_{xy}(H) = \rho_{xy}(H)/\mu_0 H$. The parameter $\mu(H)$ can be viewed as the total 'mobility' of a two-carrier system and its magnetic field dependence can be directly obtained from the experimental data of $\rho_{xy}(H)$ and $\rho_{xx}(H)$ using Eq. (3).

Eq.(1) and Eq.(2) can be rewritten as:

$$\rho_{xx}(H) = \rho_\infty \frac{\mu^2(\mu_0 H)^2}{1 + \mu^2(\mu_0 H)^2} + \rho_{xx}(0) \ , \tag{4}$$

$$R_{xy}(H) = \frac{R_0 + R_\infty \mu^2(\mu_0 H)^2}{1 + \mu^2(\mu_0 H)^2} \ , \tag{5}$$

where

$$R_\infty = \frac{R_e R_h}{R_e + R_h} \ , \tag{6}$$

$$R_0 = \frac{R_h \rho_e^2 + R_e \rho_h^2}{(\rho_e + \rho_h)^2} \quad , \tag{7}$$

$$\rho_\infty = \frac{(R_e \rho_h - R_h \rho_e)^2}{(\rho_e + \rho_h)(R_e + R_h)^2} \quad , \tag{8}$$

$$\rho_{xx}(0) = \frac{\rho_e \rho_h}{\rho_e + \rho_h} \quad . \tag{9}$$

The parameters $\rho_{xx}(0)$ and $R_0$ are the electrical resistivity at zero temperature and the Hall coefficient $R_{xy}$ in the limit of $H \to 0$ respectively, which can be directly obtained from the experimental data. After determining the parameter $\mu(H)$, then one can extract the parameters of $R_\infty(H)$ and $\rho_\infty(H)$ following Eq.(4) and Eq.(5). By substituting the derived parameters $R_\infty$, $\rho_\infty$, $R_0$ and $\rho_{xx}(0)$ into Eqs. (6)-(9) and numerically solving the equations, the field dependence of $R_e$, $R_h$, $\rho_e$ and $\rho_h$ can be uniquely determined, as shown in Fig. 4 for $BaFe_2As_2$.

**Acknowledgements** We thank M. B. Salamon, Q. M. Si, J. X. Zhu, T. Xiang, Z. Y. Weng, Z. Y. Lu, F. C. Zhang, W. Q. Chen and J. H. Dai for helpful discussions. This work was supported by the National Science Foundation of China, the National Basic Research Program of China (973 program), the PCSIRT of the Ministry of Education of China, the Fundamental Research Funds for the Central Universities of China, Zhejiang Provincial Natural Science Foundation of China, the DOE BES program "Science in 100 T" and the NHMFL-UCGP. Work at NHMFL-LANL is performed under the auspices of the National Science Foundation, Department of Energy and State of Florida.

## Figure legends:

**Figure1: The in-plane magnetoresistivity of $BaFe_2As_2$: (a) H//c and (b) H//ab.** The main figure shows the normalized magnetoresistivity, $\Delta\rho_{xx}/\rho_0$, at various temperatures. As an example, the insets plot the electrical resistivity versus magnetic field at some selected temperatures. The magnetoresistivity is nearly zero in the paramagnetic state but increases remarkably with decreasing temperature in the magnetically ordered state. Longitudinal resistivity and transverse Hall resistivity were simultaneously measured with a typical 5-probe method in pulsed fields of up to 55 T at Los Alamos National High Magnetic Field Laboratory (see Ref. [33]). Single crystals with high quality were prepared by FeAs self-flux method[10, 11].

**Figure 2: Temperature dependence of the in-plane electrical resistivity at various magnetic fields for $BaFe_2As_2$ and $Ba(Fe_{0.95}Co_{0.05})_2As_2$.** The top and bottom panels are for fields along the c-axis and the ab-plane, respectively. The zero field resistivity (see the solid line) was measured with Lakeshore 370 ac resistance bridge while cooling down the sample to the base temperature, from which an anomaly is observed at $T_M \approx 142$ K and 77K for $BaFe_2As_2$ and $Ba(Fe_{0.95}Co_{0.05})_2As_2$, respectively. The insets show the temperature dependence of the magnetoresistivity, $\Delta\rho_{xx}/\rho_0$, which vanishes exactly around $T_M$. In both compounds, the transition at $T_M$ is nearly independent of the magnetic field up to 55T, suggesting a localized nature of the magnetic order. A metal-insulator-like crossover is observed around 20T in $BaFe_2As_2$. The magetoresistivity is largely reduced in the underdoped compound $Ba(Fe_{0.95}Co_{0.05})_2As_2$, which resistivity increases with decreasing temperature below $T_M$, followed by a superconducting transition at $T_c \approx 9.5$K.

**Figure 3: The Hall resistivity versus magnetic field for (a) $BaFe_2As_2$ and (b) $Ba(Fe_{0.95}Co_{0.05})_2As_2$.** The Hall resistivity of $BaFe_2As_2$ follows linear field

dependence above $T_M$, but shows a parabolic-like behaviour in the magnetic state. It is noted that such behaviour is observed in samples coming from different groups and it becomes more pronounced in samples with a higher RRR ratio (ratio of room temperature resistivity to residual resistivity). In the doped compound, the Hall resistivity monotonically decreases with increasing magnetic field, but deviates from the linear field dependence as well. Note that the sharp decrease of $\rho_{xy}(\mu_0 H)$ at 1.5K corresponds to the superconducting transition. The inset shows the temperature dependence of the Hall coefficient $R_H$ for $BaFe_2As_2$, in which the data marked with empty triangles are from Ref. [24], derived in a dc magnetic field up to 8T.

**Figure 4: The magnetic field dependence of the derived parameters of $R_e$, $R_h$, $\rho_e$ and $\rho_h$ for $BaFe_2As_2$.** The filled and open symbols represent contributions from electrons and holes, respectively. Different symbols mark different temperatures. One can see that the Hall coefficient $R_h$ increases with magnetic field, but the absolute value of $R_e$ decreases rapidly upon applying a magnetic field, and then eventually gets saturated. The electrical resistivity contributed by hole- pocket and electron- pocket shows opposite field dependence.

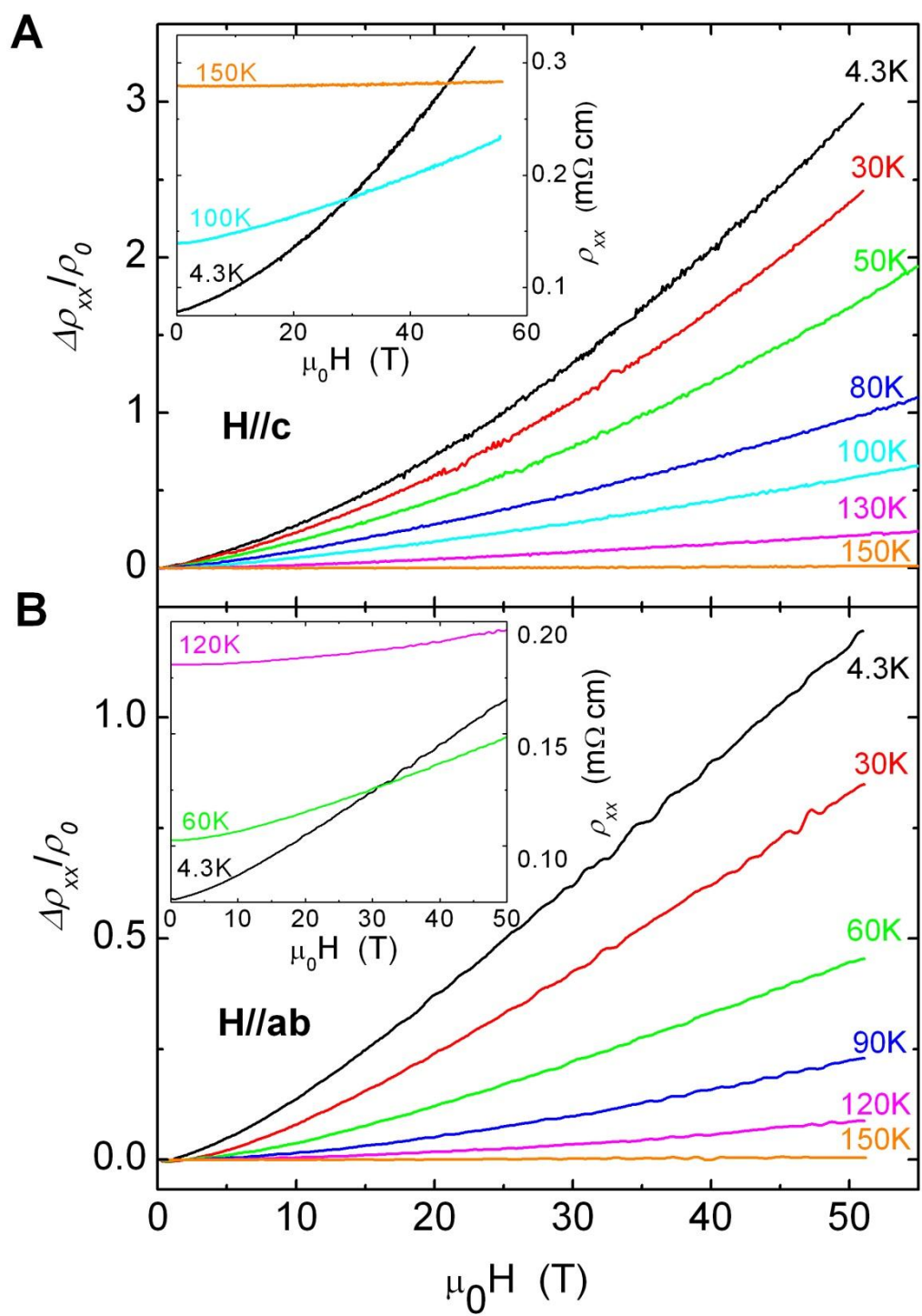

**Figure. 1 Yuan**

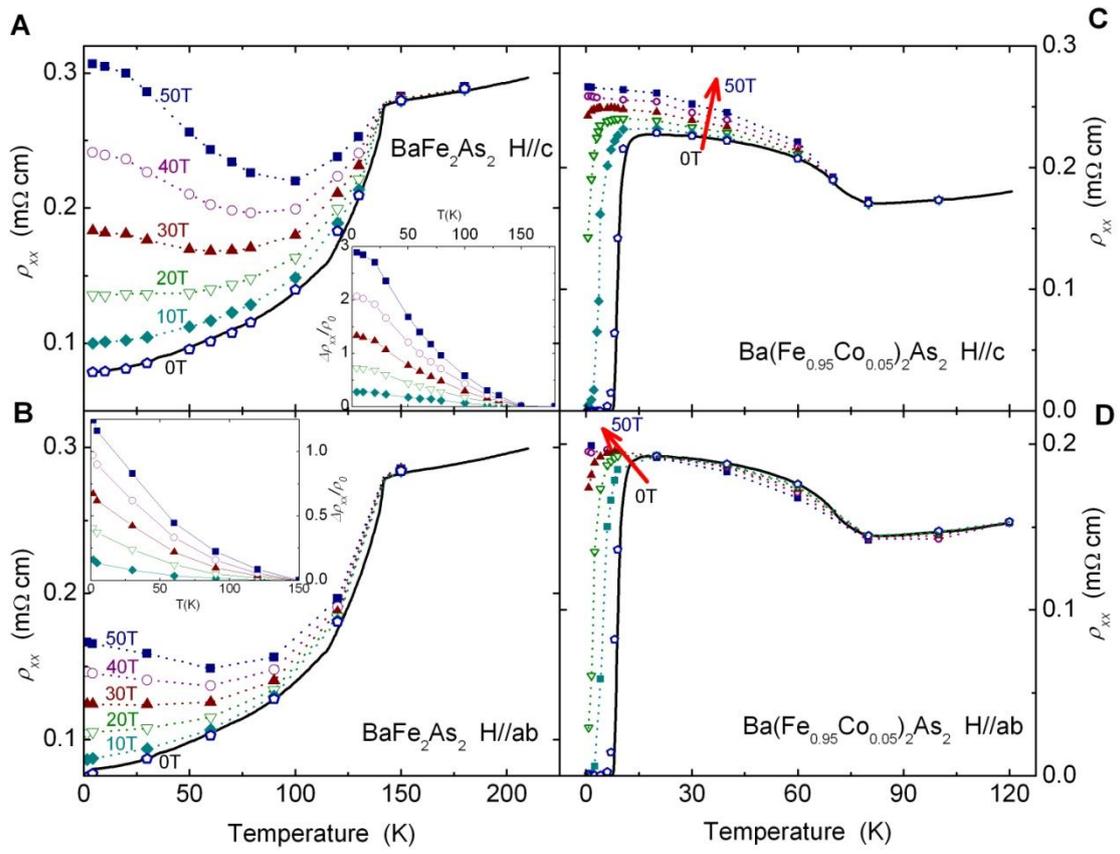

**Figure. 2 Yuan**

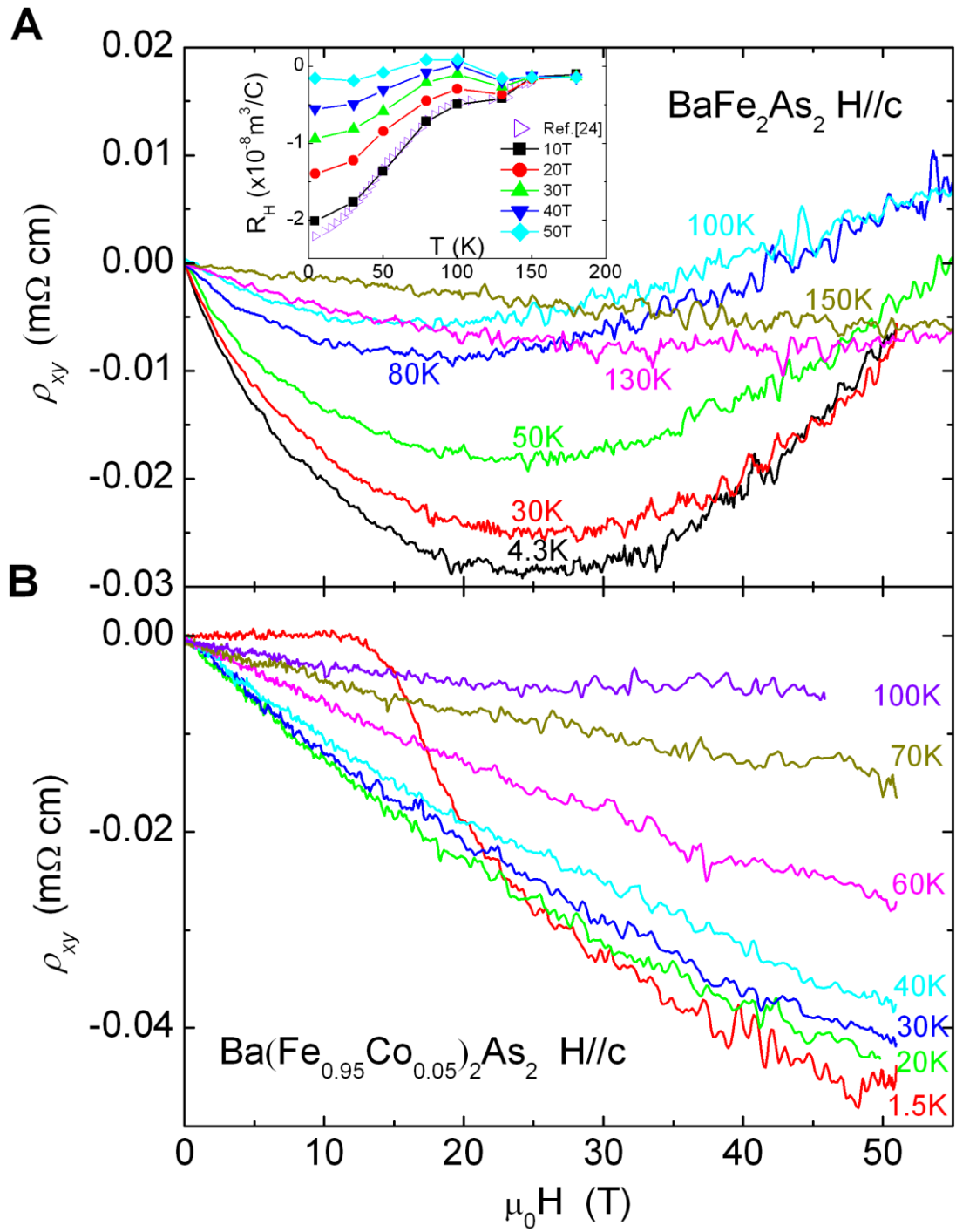

**Figure. 3 Yuan**

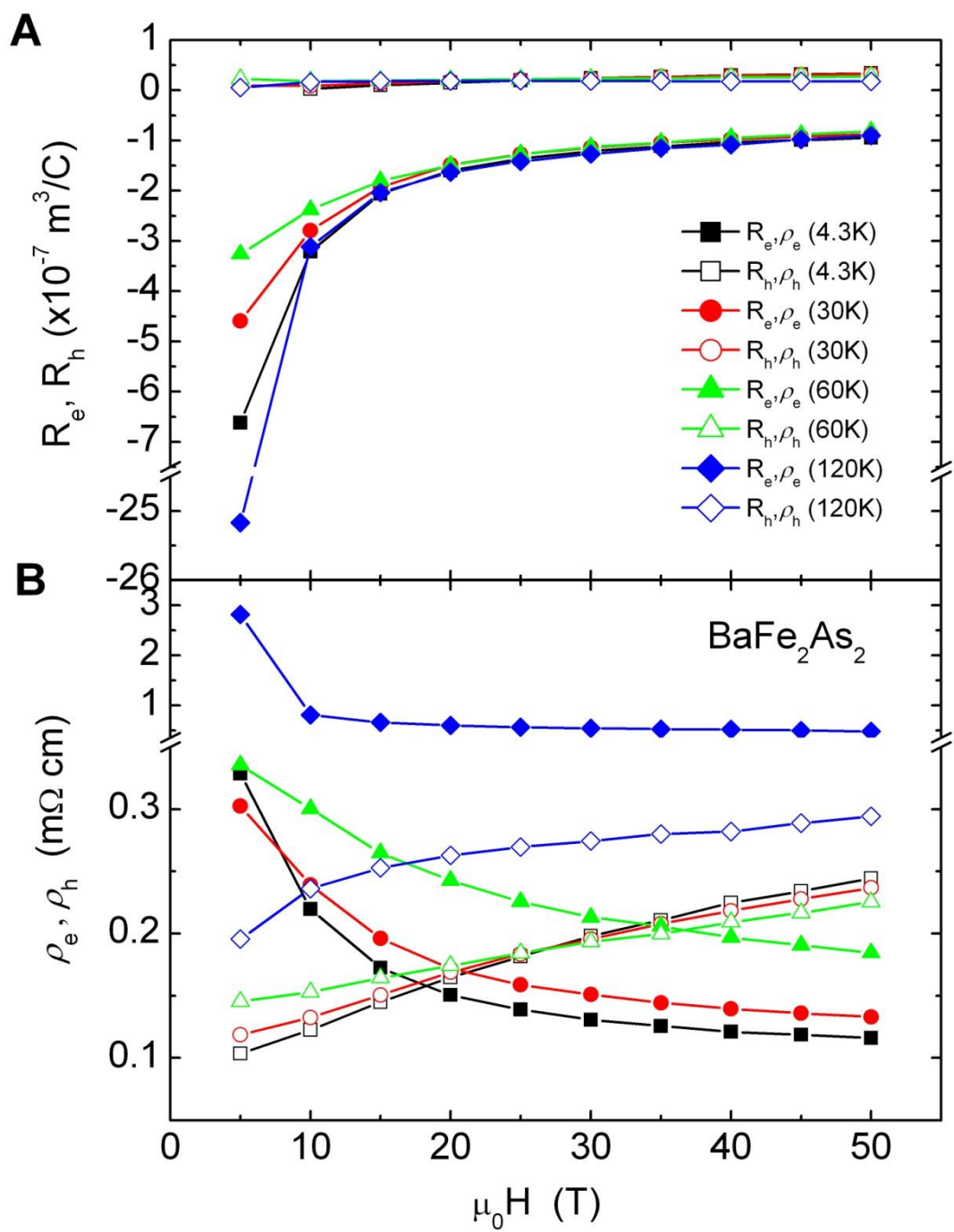

**Figure. 4 Yuan**